\documentclass[twocolumn,showpacs,preprintnumbers,amsmath,amssymb,prl,unsortedaddress,superscriptaddress]{revtex4}
\usepackage[pdftex]{graphicx}
\usepackage{dcolumn}         % Align table columns on decimal point
\usepackage{bm}              % bold math

\begin{document}

\title{Electronic energy level alignment at metal-molecule interfaces \\ with a GW approach}

\author{Isaac Tamblyn} 
  \email{itamblyn@lbl.gov}
  \affiliation{Molecular Foundry, Lawrence Berkeley National Laboratory}
\author{Pierre Darancet}
  \affiliation{Molecular Foundry, Lawrence Berkeley National Laboratory}
\author{Su Ying Quek} 
  \affiliation{Institute of High Performance Computing, Singapore}
\author{Stanimir A. Bonev} 
  \affiliation{Lawrence Livermore National Laboratory \\
   Department of Physics, Dalhousie University, Halifax, NS}

\author{Jeffrey B. Neaton} 
	\email{jbneaton@lbl.gov}
	\affiliation{Molecular Foundry, Lawrence Berkeley National Laboratory}

\date{\today}

\begin{abstract}

Using density functional theory and many-body perturbation theory within a GW approximation, we calculate the electronic structure of a metal-molecule interface consisting of benzene diamine (BDA) adsorbed on Au(111). Through direct comparison with photoemission data, we show that a conventional G$_0$W$_0$ approach can underestimate the energy of the adsorbed molecular resonance relative to the Au Fermi level by up to 0.8 eV. The source of this discrepancy is twofold: a 0.7 eV underestimate of the gas phase ionization energy (IE), and a 0.2 eV overestimate of the Au work function. Refinements to self-energy calculations within the GW framework that account for deviations in both the Au work function and BDA gas-phase IE can result in an interfacial electronic level alignment in quantitative agreement with experiment.
\end{abstract}


\maketitle

There is considerable interest in using organic materials as components in nanoscale energy conversion applications, and thus a critical need has emerged for improved knowledge and control of the electronic structure of metal-molecule interfaces. In particular, understanding how molecular addition/removal energies (ionization energy, IE; and electron affinity, EA) are altered at a metal contact is fundamental to molecular-scale transport~\cite{quek_nanolett_2007, toher_prl_2005, poulsen_naturenat_2009}, energy conversion in organic photovoltaics~\cite{scharber_advm_2006,lloyd_mt_2007}, and photo- and electrocatalytic systems~\cite{bard_science_1980}.

Understanding metal-molecule interface electronic structure with spectroscopic accuracy poses significant challenges to standard first-principles approaches. Important physical factors influencing electronic level alignment include the magnitude of the interface dipole formed upon adsorption, molecular level broadening via hybridization with substrate states, and surface polarization effects on electron addition and removal energetics. While density functional theory (DFT) approaches within standard local and semi-local approximations can often describe interface dipoles~\cite{dellangela_nanolett_2010,heimel_prl_2006,magid_jpcc_2008,biller_submitted}, hybridization, and work functions with good accuracy, prior studies~\cite{neaton_prl_2006, garcia-lastra_prb_2009, thygesen_prl_2009, freysoldt_prl_2009, rostgaard_prb_2010, chen_jcp_2010, rangel_prb_2011} have established that the impact of substrate polarization, a non-local correlation effect, is absent from mean-field Kohn-Sham states. Self-energy corrections within the GW approximation can capture this effect, with a significant impact on gaps of adsorbed molecules ($>$1 eV for small aromatic molecules). GW methods can also significantly improve the IE and EA of gas-phase molecules~\cite{blase_prb_2011, rostgaard_prb_2010} compared to canonical semi-local Kohn-Sham DFT, where the energy difference between highest occupied molecular orbital (HOMO) and lowest unoccupied molecular orbital (LUMO) energies is underestimated relative to the fundamental gap (i.e. IE - EA), even for the hypothetical ``exact'' exchange-correlation potential~\cite{perdew_prl_1983, sham_prl_1983, kummel_rmp_2008}. DFT frontier orbital energy differences can, however, provide accurate fundamental gaps if a judicious approximation within a generalized Kohn-Sham framework is used~\cite{stein_prl_2010}.

\begin{figure}[!t] 
\begin{center}
\includegraphics[width=0.5\textwidth,clip]{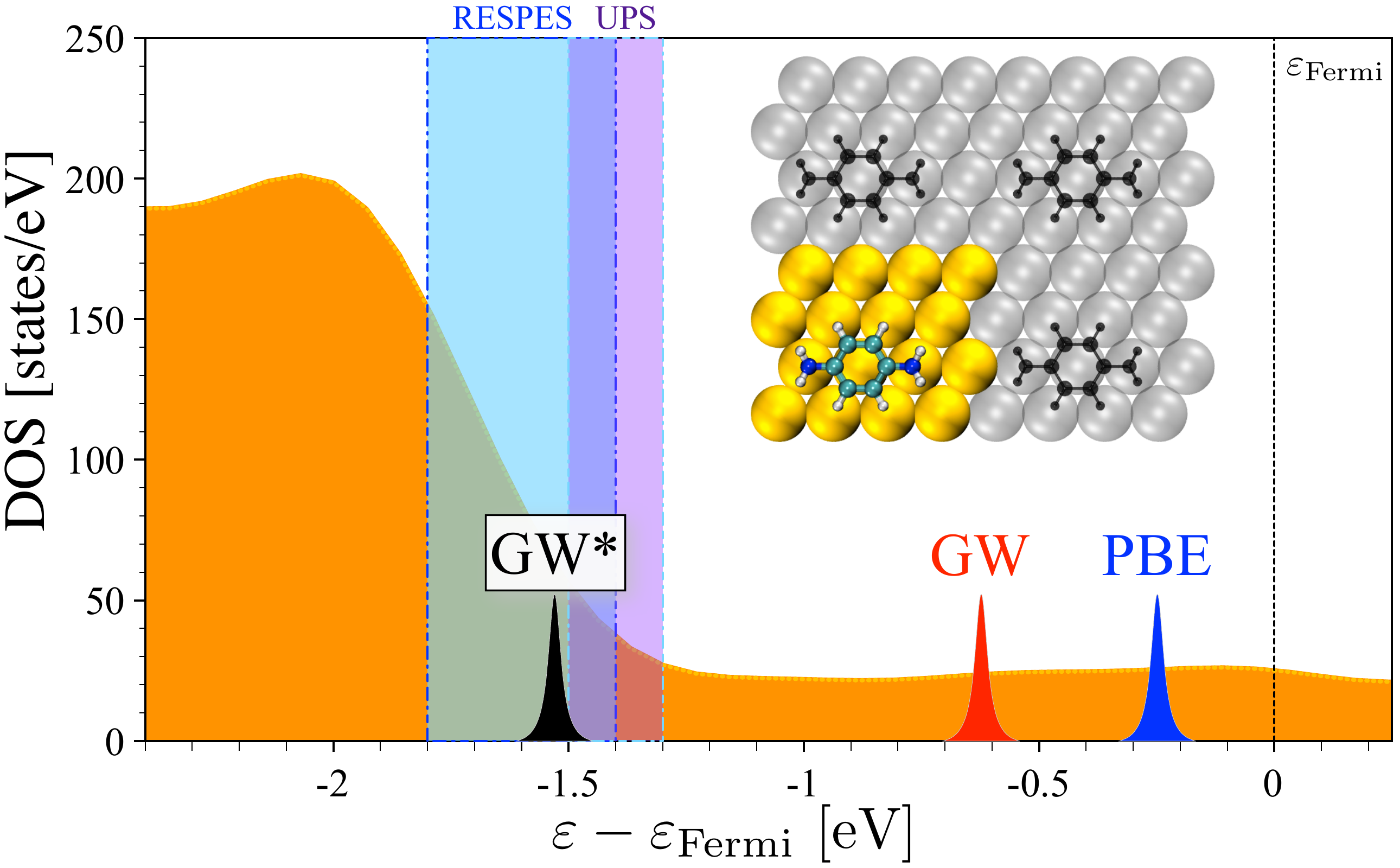}
\end{center}
\caption{\label{f:gw_dos} Electronic density of states (DOS) of 1,4-benzenediamine (BDA) above the Au(111) atop site. The inset shows geometry of the 4x4 supercell in color and periodic images in gray. The position of the highest occupied molecular orbital within DFT is -0.25~eV below the Fermi level. G$_0$W$_0$ (this work) differs from the experimental measurement by 0.8 eV. The source of this discrepancy is twofold: a 0.7 eV underestimate of the gas phase ionization energy (IE), and a 0.2 eV overestimate of the Au work function. Accounting for these errors with a \emph{post-hoc} correction, i.e. a rigid shift, produces a value (GW$^*$) in excellent agreement with experiment. Results from resonant and ultraviolet photoemission spectroscopy (RESPES/UPS)~\cite{dellangela_nanolett_2010} and their uncertainties are indicated by the shaded regions.}
\end{figure}

Previous GW calculations of energy level alignment at interfaces~\cite{rignanese_prl_2001,shaltaf_prl_2008} suggest significant improvement over DFT-GGA. In this work, we calculate the energy level alignment ($\varepsilon_{\textrm{HOMO}}$ - $\varepsilon_{\textrm{Fermi}}$, see Fig. ~\ref{f:gw_dos}) at a prototype metal-molecule interface, benzenediamine (BDA) on Au(111), comparing our GW calculations directly with photoemission spectroscopy (PES) measurements. We find that modest inaccuracies in the constituent gas-phase BDA IE and Au(111) work function within a standard G$_0$W$_0$ approach -- using a plane-wave basis set and plasmon-pole models, and requiring sums over unoccupied single-particle states -- are additive for the adsorbate system in this case, leading to a discrepancy with measurements for the HOMO resonance of up to 0.8 eV. Refinements of self-energies within the GW framework that ameliorate deviations for the isolated constituents can lead to a predicted HOMO resonance in agreement with photoemission spectroscopy. 

GW calculations of metal-molecule interfaces pose several computational challenges. First, accurate evaluation of the Fock exchange requires explicit treatment of the semicore electrons, imposing the correct nodal structure on \emph{d} states but leading to higher cutoffs and a need to treat more electrons~\cite{marini_prl_2001}. Second, metals often require a dense k-point sampling and the relevance of plasmon pole approximations can be questionable~\cite{marini_prl_2001,shih_prl_2010}. Third, for a hybrid interface comprised of a molecule (with localized states) and a metal (with delocalized states), self-energies for the constituent systems are of very different magnitudes, and taking the Kohn-Sham eigenstates as the quasiparticle wavefunctions may no longer be a good approximation~\cite{pulci_prb_1999}. Fourth, for level alignment between states of disparate character, we require absolute convergence, and, in this case, since Au self-energy corrections converge differently than those for BDA states, a large number of unoccupied states (N$_{\textrm{c}}$), together with a good extrapolation scheme~\cite{samsonidze_prl_2011,static_remainder}, can be necessary. This is a particular challenge given the concomitant need for a large supercell. 

Our GW calculations are performed using the BerkeleyGW~\cite{BerkeleyGW} and Abinit~\cite{abinit} codes, following a well-established G$_0$W$_0$ approach ~\cite{hybertsen_prb_1986}. Equilibrium geometries of molecular BDA in the gas-phase and physisorbed on Au(111) are determined using DFT~\cite{paratec} within the PBE~\cite{pbe} generalized gradient approximation (GGA). The molecule is flat relative to the surface, at a height of 3.5 \AA\ above the topmost layer of Au. This is consistent with relaxed geometries obtained using a van der Waals corrected density functional~\cite{li_prep_2011}. Norm-conserving pseudopotentials [20] are used with a plane- wave basis (60 Ry cutoff) for structural relaxations and includes 5\emph{s} and 5\emph{p} semi-core states for Au. The surface is modeled with a 4x4 supercell containing 4 layers of Au (roughly 9$\times k_{\textrm{Fermi}}^{-1}$, where $k_{\textrm{Fermi}}$ is the Thomas-Fermi wavevector of gold), a single BDA molecule (see inset Figure~\ref{f:gw_dos}), and the equivalent of 10 layers of vacuum.  The metal work function and magnitude of $|\varepsilon_{\textrm{HOMO}} - \varepsilon_{\textrm{Fermi}}|$ change less than 0.05 eV in DFT when the depth of the slab is doubled. The theoretical in-plane bulk lattice parameter is used for Au (a=4.18~\AA). The supercell Brillouin zone (BZ) is sampled using a 4x4x1 \textbf{k}-grid. Gas-phase BDA is modeled using the same supercell, in the absence of Au, and using a Coulomb truncation.

For all calculations, 6 Ry planewave expansion cutoff is used for the dielectric function, which, for the majority of our work, is extended to finite frequencies with a generalized plasmon pole (GPP) model~\cite{hybertsen_prb_1986}. Doubling this cutoff and the number of unoccupied states used in constructing the response function results in negligible changes to BDA gas-phase IE, $\sim$0.15 eV. Updates to \emph{G} and \emph{W} use quasiparticle energies from the previous cycle and a linear fit to a coarse sampling of self-energy corrections to high energy states ($\varepsilon > $ 6~eV above vacuum). The number of states used to construct $\epsilon^{-1}$ was held fixed at 2048 bands. For BDA-Au(111) interface calculations, our sum over the unoccupied subspace includes more than 1400 conduction bands (30 eV above $\varepsilon_{\textrm{Fermi}}$), a number which, as we show, still falls considerably short of convergence. Gas-phase results are based on a sum of over 5100 conduction bands ($\sim$ 80~eV above vacuum) for the Coulomb-Hole term. For calculations of bulk Au, we compare with results from the Godby-Needs~\cite{godby_prl_1989} GPP model, as well as a an explicit evaluation of $\epsilon^{-1} (\textbf{q},\omega)$ for more than 200 frequencies up to about 100 eV~\cite{lebegue_prb_2003}. For bulk Au, the BZ is sampled with a 14$^3$ \textbf{k}-point grid, and 500 conduction bands is used ($\sim$ 600 eV above $\varepsilon_{\textrm{Fermi}}$). 

Table~\ref{t:gas_phase} summarizes our results for the IE and EA of gas-phase BDA at different levels of theory. Relative to measured photoemission, DFT-GGA underestimates the IE by over 3 eV, consistent with previous work~\cite{quek_nanolett_2007}. Our GW calculations are a significant improvement over DFT-PBE, within 0.7 eV or better of experiment (after extrapolating the unoccupied states to infinity), depending on whether $G$ or $W$ is updated. The error relative to experiment for G$_1$W$_1$ is 0.5 eV, or just 7\%. Remarkably, with just 1000 unoccupied states and no extrapolation, the IE is about 1 eV smaller, illustrating the slow convergence of the IE with respect to unoccupied states. Doubling the number of unoccupied states reduces the IE by only 0.2~eV. Use of an extrapolation scheme for this slow convergence, associated with the Coloumb-Hole term of the GW self-energy, is therefore crucial when comparing GW to experiment. Extrapolations $N_{\textrm{c}}\rightarrow \infty$ are determined by fitting the Coulomb-Hole term to the form $a + N_{\textrm{c}}^{-\frac{1}{x^*}}$, where $x^*$ is determined from a similar fit to the \emph{static} Coulomb hole screened-exchange (COHSEX) approximation (using the same set of convergence parameters). Within static COHSEX, the asymptote $a$ does not directly depend on $N_{\textrm{c}}$, allowing for clean optimization of the exponent. Encouragingly, our scheme seems consistent with results computed using a completion method recently proposed by Deslippe et al.~\cite{static_remainder}.

\begin{table}[tb]
\begin{center}
  \includegraphics[width=0.5\textwidth,clip]{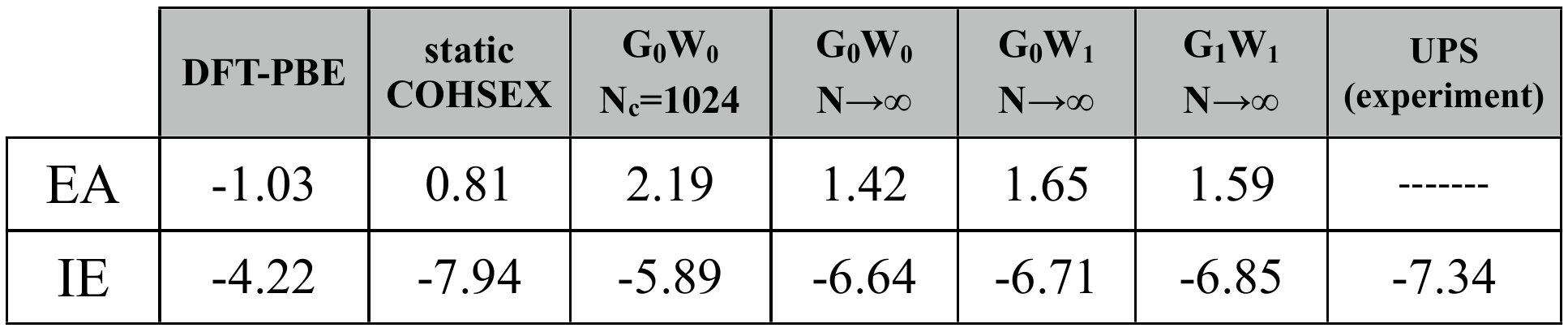}
\end{center}
\caption{\label{t:gas_phase} Energy levels for the frontier molecular orbitals in the gas phase calculated using different electronic structure methods. All values are in eV. Results for DFT-PBE correspond to Kohn-Sham eigenvalues of the neutral molecule. Values for $N_{\textrm{c}}\rightarrow \infty$ are based on an extrapolation of the Coulomb-Hole term (see text). Uncertainties in the experimental~\cite{streets_chl_1972, cabelli_jacs_1981} value of the vertical IE is $\pm$ 0.03 eV.}
\end{table}

For bulk Au, we find that the GPP models of Hybertsen-Louie and Godby-Needs leave the DFT-PBE bandwidth relatively unchanged, in adequate agreement with XPS measurements~\cite{smith_prb_1974} (overestimate of 12\% and 9\% respectively). Using a fully frequency dependent dielectric function results in a 6\% overestimate relative to experiment. Better agreement will likely require going beyond the RPA~\cite{mahan_prl_1989, northrup_prb_1989}. Interestingly, DFT-PBE provides the best Au(111) work function~\cite{work_function_comment}, 5.2 eV (within 0.05 eV of experiment~\cite{hansson_prb_1978}). With either of the GPP values used here, the Au(111) work function is larger than experiment by 0.5 eV, and by 0.2 eV with a fully frequency dependent dielectric function. For comparison, Faleev et al.~\cite{faleev_prb_2010} report a G$_0$W$_0$ (and numerical $\epsilon^{-1}$) work function for Al(111) which is 0.06 eV too \emph{small} compared to experiment (4.18 eV vs 4.24 $\pm$~0.02 eV), with a self-consistent GW approach producing nearly identical values.

To identify the HOMO energy of the BDA adsorbate, we project the DFT Kohn-Sham Hamiltonian, $\hat H_{\kappa}$, of the BDA-Au(111) supercell onto the DFT-PBE orbitals calculated from the isolated molecule, $|\iota \rangle$. For BDA physisorbed on Au(111), $|\langle \iota_{\textrm{HOMO}} | \kappa_i \rangle|^{2}$ $\sim$ 0.9 at the $\Gamma$ point, indicating such a projection is spectroscopically meaningful in this case. We note that as the BDA HOMO resonance is not the highest occupied state in our calculation, in general we would not expect its value to agree with UPS, at least within the semi-local KS framework used here. Indeed, evaluating $\langle \iota_{HOMO} | \hat H_{\kappa} | \iota_{HOMO} \rangle$, we obtain -0.25 eV, a resonance value too shallow compared to photoemission, which place the HOMO -1.4~$\pm$~0.1 eV below $\varepsilon_{\textrm{Fermi}}$.  

To evaluate GW self-energy corrections for the BDA adsorbate HOMO energy, we follow the above approach and evaluate the matrix element $\langle \iota_{\textrm{HOMO}}|\hat{\Sigma}_{\kappa}|\iota_{\textrm{HOMO}}\rangle$. Since $\hat{\Sigma}_{\kappa}$ is approximately diagonal in the molecular basis $\{| \iota_{\rm{j}} \rangle\}$, this approach is approximately equivalent to evaluating diagonal and off-diagonal elements of the full energy-dependent self-energy matrix, diagonalizing, and projecting onto the surface states to identify the adsorbate HOMO resonance energy measured spectroscopically~\cite{pulci_prb_1999, rangel_prb_2011}. This workaround, with its substantially-reduced computational cost, is strictly valid in a ``weak-coupling" limit, where both $\hat H_{\kappa}$ and $\hat{\Sigma}_{\kappa}$ are diagonal in the basis of gas-phase orbitals, as is the case here.

\begin{figure}[!t]
\begin{center}
\includegraphics[width=0.5\textwidth,clip]{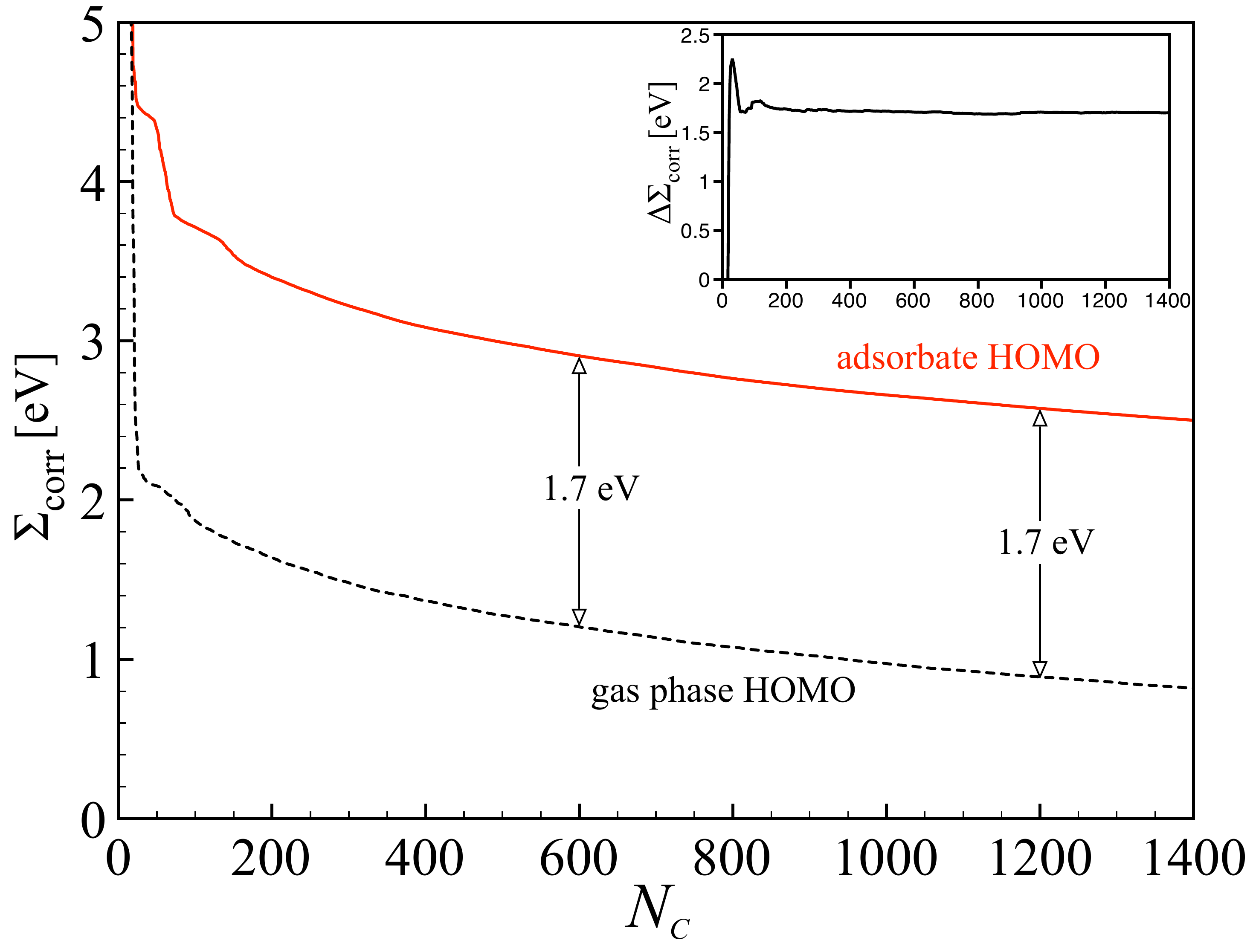}
\end{center}
\caption{\label{f:correlation_convergence} $\Sigma_{\textrm{corr}}$ for the isolated (dashed line) molecule and molecular resonance (solid line) in the monolayer as a function of the number of conduction bands used in the evaluation of $\hat{\Sigma}$. Although the absolute value of these terms converge slowly, by $N_c = $ 600 their difference, $\Delta \Sigma_{\textrm{corr}}$ (inset), has stabilized to 1.7 eV, attributable to non-local static correlations from the metal substrate. The nonlinear behavior of $\Sigma_{\textrm{corr}}(N_{\textrm{c}})$ for $N_{\textrm{c}} < 200$ reflects the character of low energy conduction bands, which are highly system specific.}
\end{figure}

To understand the adsorbate result relative to the gas-phase, we follow previous work~\cite{neaton_prl_2006} and partition the self-energy correction into two contributions: the Fock exchange, $\hat{\Sigma}_{\textrm{\textrm{X}}}$; and the portion containing static and dynamical correlation, $\hat{\Sigma}_{\textrm{corr}}$. We find that $\Sigma_{\textrm{X}}$ for the adsorbate HOMO differs that obtained for gas-phase BDA, $\Sigma_{\textrm{X}}(\textrm{adsorbate}) - \Sigma_{\textrm{X}}(\textrm{gas phase})= \textrm{0.4 eV}$, which can be understood in terms of the non-zero overlap of the BDA HOMO with the Au wavefunctions. Unlike $\hat \Sigma_{\textrm{X}}$, $\hat \Sigma_{\textrm{corr}}$ involves a difficult-to-converge sum over the unoccupied space (as with gas-phase BDA). However, from Fig.~\ref{f:correlation_convergence}, the \emph{difference} between $\Sigma_{\textrm{corr}}$ of the isolated molecule and the adsorbate monolayer converges much faster, with a modest sum of 600 bands (Fig. ~\ref{f:correlation_convergence}) to 1.7 eV. As with benzene on graphite~\cite{neaton_prl_2006}, this response is due almost entirely to static polarization: an electrostatic image charge model (with a calculated~\cite{lam_jpcm_1993} image plane of 1.47 \AA\ above the Au surface) predicts a value of $\Delta \Sigma_{\textrm{corr}} = 1.8$ eV. 

As shown in Fig.~\ref{f:gw_dos}, our G$_0$W$_0$ corrections, in the limit of $N_{\textrm{c}}\rightarrow \infty$, lead to a BDA adsorbate HOMO energy of -0.64 eV, a significant underestimate of the experimental value of -1.4 eV. Given the converged self-energy corrections for the isolated molecule, the metal work function, and the change in the correlation energy upon adsorption, we can understand the disagreement between theory and experiment for the adsorbate system as originating with the underestimate of the gas-phase BDA IE and the overestimate of the Au(111) work function. If we account for the discrepancies of the \emph{isolated} systems with a rigid shift (GW$^*$ in Fig.~\ref{f:gw_dos}), good agreement between theory and experiment for the \emph{composite} system is obtained.

Our results illustrate that the accuracy of energy level alignment at a metal-molecule interface with a given GW approach is limited by its ability to describe the IE of an isolated molecule and the metal work function. In a stronger-coupling limit, dynamical contributions to electrode polarization would become important, as has been noted before~\cite{neaton_prl_2006, thygesen_prl_2009, freysoldt_prl_2009}. In such a case, classical static polarization models are less valid, and $\Sigma_{\textrm{corr}}$ obtained from GW must be used. Furthermore, $\Sigma_{\kappa}$ may no longer be diagonal in the basis of gas phase orbitals, necessitating full evaluation of the self-energy operator.

In conclusion, through direct comparison with photoemission spectroscopy, we have demonstrated the advantages and limitations of an existing G$_0$W$_0$ approach in describing the electronic structure of a molecule adsorbed to a metal substrate. GW improves upon PBE, particularly in its inclusion of nonlocal correlation, but is limited, at least in the approach considered here, by its ability to predict gas-phase IEs and work functions. Computationally-tractable refinements that improve accuracy of these quantities will result in better quantitative agreement for level alignment.

\section{Acknowledgments}
The authors acknowledge fruitful discussions with Prof. L. Kronik, and J. Deslippe, P. Doak, G. Samsonidze, D. Strubbe, S. Sharifzadeh, and A. Zayak. Work at the Molecular Foundry was supported by the Office of Science, Office of Basic Energy Sciences, of the U.S. Department of Energy under Contract No. DEAC02-05CH11231. Computer simulations were performed using NERSC-Hopper and ACE-net. I.T. acknowledges financial support from NSERC. 

%\bibliography{references} 

\begin{thebibliography}{46}
\bibitem[{wor()}]{work_function_comment}
\bibinfo{note}{Our GW work function of gold is obtained by calculating the
  Fermi level offset (relative to PBE) from a fully converged G$_0$W$_0$
  calculation of bulk fcc gold, and then aligning it by matching the z-averaged
  $V_{\textrm{Hartree}}$ of the bulk with the bulk-like portion deep within the
  slab.}
\end{thebibliography}

\end{document}